**A Bibliometric Approach to Tracking International Scientific Migration**

Henk F. Moed, Informetric Research Group, Elsevier | Radarweg 29, 1043 NX Amsterdam, The Netherlands | Email: h.moed@elsevier.com | Phone: +31 20 485 3436

Gali Halevi, Informetric Research Group, Elsevier | 360 Park Av. South, New York NY 10011| Email: g.halevi@elsevier.com | Phone: +1 646 248 9464

**Abstract**

A bibliometric approach is explored to tracking international scientific migration, based on an analysis of the affiliation countries of authors publishing in peer reviewed journals indexed in Scopus™. The paper introduces a model that relates base concepts in the study of migration to bibliometric constructs, and discusses the potentialities and limitations of a bibliometric approach both with respect to data accuracy and interpretation. Synchronous and asynchronous analyses are presented for 10 rapidly growing countries and 7 scientifically established countries. Rough error rates of the proposed indicators are estimated. It is concluded that the bibliometric approach is promising provided that its outcomes are interpreted with care, based on insight into the limits and potentialities of the bibliometric approach, and combined with complementary data, obtained, for instance, from researchers' Curricula Vitae or survey or questionnaire based data. Error rates for units of assessment with indicator values based on sufficiently large numbers are estimated to be fairly below 10 per cent, but can be expected to vary substantially among countries of origin, especially between Asian countries and Western countries.

## 1. Introduction

Scientific networks, collaboration and exchange have been the centre of attention in numerous research articles and conferences' discussions. The main reason for the increased interest in these topics has been the premise that these types of exchanges benefit scientific progress in that they foster innovation, enhance and enable the flow of ideas between scientists in different institutions (Baruffaldi and Landoni, 2012; Biondo, 2012; Di Maria and Lazarova, 2012). In addition to the actual growth of science and scientific activity, there has been much effort to show that such progress benefits the economy through a line of investigation linking basic research to patents production. Bibliometrics took a main methodological role in the studies of co-authorship and its results as indicators of collaborative trends (Abramo et al., 2012; Gutiérrez-Vela et al., 2012, Snaith, 2012, Yang and Tang, 2012) by using affiliation information embedded in the bibliographic data of publications. In addition to the ability to track and sketch scientific collaborations between institutions, the availability of author profiles and their affiliation information in Scopus™ has also made possible the tracking of scientific migration from country to country. One of the first studies exploring the bibliometric approach to the study of scientific migration was published by Laudel (2003). Despite its limitations which will be discussed later on, such information is of immense value to the ability to study research migration and use it as a way to inform science policy and track the formation of research excellence centres as they draw domestic and international talent to their doors.

Scientific migration or mobility, although related to the formation of networks and collaboration, has unique characteristics and far reaching implications that go beyond the development of collaborative scientific activities. It can be the result of international collaboration, or lead to new collaboration ties with foreign institutions, but it has a much wider background and impact. In addition to its impact on immigration rates, economy and culture, research migration has professional implications. Enhanced scientific contribution to the receiving country, the enrichment of its scientific strength, the flow is new ideas and perspectives in different areas of research as well as its potential do develop new products and technologies are potential outcomes of research migration.

This article explores the use of bibliometric data in the study of international scientific migration, by studying researchers' migration trends between 10 scientifically developing and 7 developed countries (see Table 2 in Section 3 ) and sketches some of its major trends. The current paper is a follow-up of earlier articles published by the authors and their colleagues (Plume, 2012a, Plume, 2012b, Moed, Aisati and Plume, 2013).The latter paper introduces some of the base concepts of the bibliometric study of international scientific migration and presents results for a small set of countries, focusing on so called "migration balances". In the current paper, the base concepts are further developed; a bibliometric model of migration is presented and an analysis of the effects of inaccuracies in author profiling upon bibliometric indicators is presented. In addition, the set of study countries is expanded from 5 to 17.

The paper is structured as follows: Section 2 covers the model on which the international scientific migration analysis is based, and draws important consequences for the interpretation of the migration patterns of authors based on their publication profiles in Scopus. Moreover, it distinguishes two data-analytical approaches, a synchronous and an asynchronous one. Section 3 describes the data collection and presents the list of 17 countries studied in this paper. Section 4 gives an overview of the main outcomes of the exploratory study. When assessing the *accuracy* of migration counts a crucial issue is the accuracy of the author profiling routine in Scopus™. This routine aims to correct for homonyms (one author name relates to different persons - common names) and synonyms (a researcher has more than one author names – split identity). In the ideal case, each researcher publishing articles in Scopus™-covered sources has one single author ID, and each author ID relates to one single researcher only. Inaccuracies in author profiling may have direct consequences on the accuracy of the migration data. Therefore, Section 5 examines the degree of inaccuracy of Scopus™ author profiling and the implications it has for the migration data analyzed in this paper. Finally, Section 6 draws conclusions on the potentialities and limitations of the method, and formulates questions for further research.

## 2. International Scientific Migration Analysis – Model and Definitions

In order to study migration patterns we have defined a specific model for the analysis in which the move of researchers from one country to another can be more easily tracked. Since a bibliometric method is used, the connection between the theoretical construct and the bibliometric one is specified in Table 1.

| Theoretical Concept / Interpretation | Bibiometric Constructs |
|---|---|
| Researcher | Scopus Author ID |
| Being an active researcher in a particular year | Publishing an article in that year |
| Being a currently active researcher | Publishing in the current year |
| Researcher starting a scientific career during years T1-T2 | First publication appears in T1-T2 |
| Junior researcher in a particular year | First publication is a few years old |
| Researcher migrating from country A to B | Publishing author's "work" country changes over time from A to B |

Table 1: Theoretical concepts and their bibliometric constructs.

Using a bibliometric methodology, this study examined the move of researchers from one country to another via the affiliations stated on their publications through the years. A crucial point that has to be made here is that bibliometric research allows us to track mobility only to the extent that researchers publish and that their affiliation is stated on their publication in a way that can be traced back to them. In this model a country relates to the geographic location of an author's working place during the time the work described in a paper was carried out, and *not* to his/her nationality, country of birth or official country of residence. Therefore, it is important to note that any move prior to the appearance of a publication in the literature cannot be measured using this methodology (see Figure 1).

The model, as sketched in Figure 1 assumes the following career phases and their corresponding publication behavior: (1) a researchers' first publication occurs as a PhD student; (2) a researcher publishes his/her PhD dissertation; (3) further publications occur as the researcher moves to a post doc position and progresses his/her career in the institution as a senior. The manner by which researchers might move from one country to another can be outlined in several paths: (1) a researcher moves abroad to pursue a Master and PhD degree and publishes his/her first works in the receiving country. Later on s/he moves back to their origin country and continues publishing there; (2) a researcher pursues his/her Masters in the origin country, then moves to a receiving country to pursue a PhD degree where s/he publishes his/her first work, then moves back to the origin country where s/he continues to publish throughout his/her career; (3) a researchers pursues his/her master and PhD degrees in the origin country where they publish their first works, then moves to a receiving country to complete a post doctorate where they work and publish for a certain period of time and moves back to the origin country where they continue to publish. Of course there are several other scenarios and situations. The point to be made in this context is that scenarios 1 and 2 lead to the same bibliometric pattern, and therefore cannot be distinguished purely on the basis of bibliometric data.

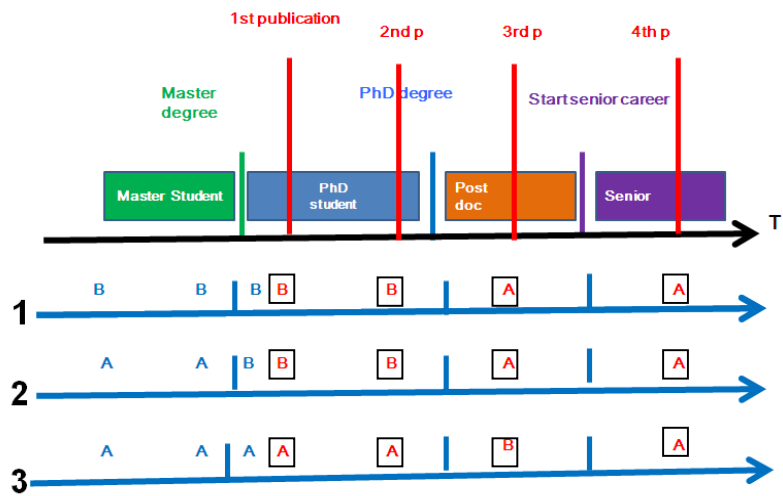

Figure 1: Three scenarios of international scientific migration and their bibliometric reflection. A and B represent the country of origin, and the receiving country, respectively. The bibliometric approach explored in this paper reveals the data included in the squares.

## 3. Data collection and analysis

We collected the research output of 17 countries among which 10 are considered growing countries (noted in italic) and 7 are considered as established from different regions in the world (see Table 2). For each country, the research output for 2000-2012 was collected.

| D8 | EU | BRIC | Other |
|---|---|---|---|
| EGYPT (EGY) | ROMANIA (ROM) | BRAZIL (BRA) | THAILAND (THA) |
| IRAN (IRN) | PORTUGAL (PRT) | CHINA (CHN) | |
| MALAYSIA (MYS) | | INDIA (IND) | AUSTRALIA (AUS) |
| PAKISTAN (PAK) | GERMANY (DEU) | | JAPAN (JPN) |

|  | ITALY (ITA) |  | UNITED STATES(USA) |
|--|-------------|--|--------------------|
|  | NETHERLANDS (NLD) |  |  |
|  | United Kingdom (GRB) |  |  |

Table 2: Countries included in the study

In order to trace the move of researchers from one country to another we used the unique Author ID offered by Scopus™ as a way to identify individual authors. In Scopus™, the affiliations associated with an author through their publications are kept and become a part of the unique author profile constructed within Scopus™. This allows for an analysis of migration because one can identify in which institution and country an author published. Moreover, the fact that the affiliation is tracked per author allows for a distinction between international migration and co-authorship patterns as separate indicators of areas of mobility versus collaboration.

Figure 2 presents the number of authors publishing at least one paper in the year 2011 for each country included in this study. This number varies from around 800,000 for China and the USA, to around 200,000 for Germany, Japan, and United Kingdom, and ranges between 14,000 and 20,000 for Portugal, Romania, Thailand, Egypt, and Pakistan.

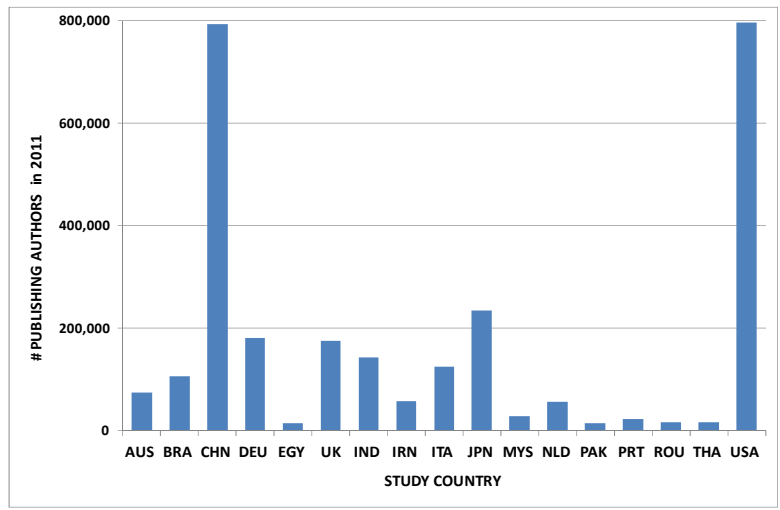

Figure 2: Number of publishing authors in 2011 per study country; AUS=Australia; BRA=Brazil; CHN=China; DEU=Germany; EGY=Egypt; UK=United Kingdom; IND=India; IRN=Iran; ITA=Italy; JPN=Japan; MYS=Malaysia; NLD=Netherlands; PAK=Pakistan; PRT=Portugal; ROU=Romania; THA=Thailand; USA=United States.

The dataset used in this study included 100,830 authors for the 17 countries studied. In order to study the rates of mobility and identify these trends per country, we devised a dataset that included specific fields (see Table 3). It must be noted that authors publishing one or more papers in one single year (so called transients) were *not* included in the analysis. Given the short duration of their publication time period, these authors are not considered as career scientists; moreover, they do not provide information on how authors move from one country to another.

| Dataset Field | Explanation |
|---|---|
| Source country<br><br>Destination- Country | These two fields include the origin country (source) and the receiving country (destination).<br><br>Example: FROM Australia TO Brazil |
| Total number of authors moving from source-country to destination country | This field aggregates the total number of authors that moved from country A to country B.<br><br>Example: The total number of authors moving FROM Australia TO Brazil = 103 |
| Total number of authors moving to destination country<br><br>Total number of authors moving from source country | These two field denote the total number of authors moving to a certain country and the total number of authors moving from a certain country<br><br>Example: the TOTAL number of authors moving TO Australia = 6,053 and the total number of authors moving FROM Brazil = 2,247 |
| % authors moving from source country to destination country, relative to total number of authors moving to Destination country | Example: The % of authors moving FROM Brazil TO Australia relative to the total number of authors moving TO Australia = 1.7 |
| % authors moving from Source country to destination country, relative to total number of authors moving from source country . This index is denoted as Relative Migration Index (RMI) in Section 4. | Example: The % of Authors moving FROM Brazil TO Australia compared to the total number of authors moving FROM Brazil = 4.6 |

Table 3: Dataset fields structure.

Data analysis approach

The data was analyzed based on two complimentary approaches; synchronous and asynchronous. In the synchronous approach we analyzed 2011 publications looking back at authors' output between 2001 and 2011. In the asynchronous approach we analyzed 2001-2011 publications looking at the authors who started their careers during 2001-2003 (see Table 4). The main reason for choosing these approaches was to allow for distinct observations of migration to and from different countries and look at rates of mobility within countries.

| Feature | Synchronous | Asynchronous |
|---|---|---|
| Publication years analyzed | FIXED (2011) | VARIABLE (2001-2011) |
| Starting years of Authors' Careers | VARIABLE (2001-2010) | FIXED (2001-2003) |

Table 4: Data Analysis: Synchronous vs. Asynchronous approach.

## 4. Results

### 4.1 Results of Synchronous Analysis

Using the synchronous approach, analyzing the 2011 publications and including authors who started their careers from 2001 to 2010 we were able to trace the degree of migration between various countries. In this approach the study countries are considered as *destination* countries. Our analysis was based on both absolute numbers of migrating authors as well as on a relative migration index (RMI) . The main reason to choosing a two-phase approach to the analysis is that calculating absolute numbers does not allow for the sheer size of the scientific network of the source country to be taken into consideration, therefore, a relative index is an appropriate method to do so. The RMI measures from the point of view of a source country A the preference that authors migrating from it have for a particular destination country B is defined as the number of authors moving from A to B divided by the total number of authors moving from A to any of the 17 destination countries analysed in the study. RMI obtains values between and including 0 and 1.

The three main scientific destinations were found to be (1) USA (2) China and (3) UK. Table 5 shows for each of these three countries the top 5 source countries in terms of number of authors moving from a source country to the destination country ("# authors"), and – based on the set of 25 source countries with the largest number of authors moving to a destination country – the top 5 source countries according to a Relative Migration Index (RMI). It demonstrates the difference between migration trends based on absolute numbers and those based on a relative index (RMI).

Table 5 shows, for instance, that based on an analysis of absolute numbers of migrating patterns to the USA, the largest scientific migration to the USA comes, in this order of rank, from (1) China (2) UK (3) Canada (4) Germany and (5) India.

| | Ranked by # Authors | | | | Ranked by RMI | | |
|---|---|---|---|---|---|---|---|
| Rank | Source country | # Authors | RMI | | Source country | # Authors | RMI |
| | | | | | | | |
| ***Destination Country: USA*** | | | | | | | |
| | | | | | | | |
| 1 | China | 6,305 | 0.66 | | Israel | 1,128 | 0.69 |
| 2 | UK | 5,373 | 0.37 | | China | 6,305 | 0.66 |
| 3 | Canada | 4,645 | 0.48 | | India | 3,307 | 0.61 |
| 4 | Germany | 4,575 | 0.41 | | Turkey | 729 | 0.60 |
| 5 | India | 3,307 | 0.61 | | Mexico | 593 | 0.58 |
| | | | | | | | |
| ***Destination Country: UK*** | | | | | | | |
| | | | | | | | |
| 1 | United States | 4,848 | 0.16 | | Ireland | 541 | 0.35 |
| 2 | Germany | 1,896 | 0.17 | | Greece | 424 | 0.32 |
| 3 | France | 1,318 | 0.14 | | South Africa | 227 | 0.26 |
| 4 | Australia | 1,259 | 0.24 | | Australia | 1259 | 0.24 |
| 5 | Italy | 1,147 | 0.21 | | Nw Zealand | 310 | 0.22 |
| | | | | | | | |
| ***Destination Country: China*** | | | | | | | |
| | | | | | | | |
| 1 | USA | 7,170 | 0.23 | | Taiwan | 1,048 | 0.48 |
| 2 | Japan | 1,606 | 0.25 | | Singapore | 843 | 0.40 |
| 3 | Taiwan | 1,048 | 0.48 | | Japan | 1,606 | 0.25 |
| 4 | UK | 1,017 | 0.07 | | USA | 7,170 | 0.23 |
| 5 | Canada | 998 | 0.10 | | South Korea | 511 | 0.15 |

Table 5: Migration to the USA, UK and China; Top 5 source countries in terms of number of authors moving from the source country to the destination country ("# authors"), and – based on the set of top 25 of countries according to # authors – the top 5 countries according to a Relative migration Index (RMI), defined as the number of authors moving from a source country A to destination country B divided by the total number of authors moving from A to any of the 17 destination countries analysed in the study.

However, based on a Relative Migration Index calculation, a different picture emerges. Compared to the total number of migrating authors, Israel is actually a leading exporter of scientists to the USA followed by China, India, Turkey and others (see the right hand side of Table 5)

Based on calculations of absolute numbers of scientists moving to the UK, the leading origin countries are USA, Germany, France, Australia. RMI calculation reveals a much different picture. Relative to the number of total migrating scientists, the leading exporters of researchers to the UK are Ireland, Greece, South Africa, Australia and New Zealand.

Finally, migration to China based on absolute numbers is showing a clear domination of the USA as the major exporter of scientists followed by Japan, Taiwan, UK and Canada. But normalizing by size of the source country, Table 5 shows the strong relationship not only with Taiwan and Japan, but also with Singapore.

4.2 Results of Asynchronous Analysis

The asynchronous approach enabled a characterization of migration patterns of study countries conceived primarily as source countriesFigure 6 compares the percentage of authors who stay within their countrywith that of authors who move permanently and that for researchers who migrate yet return to the origin country . 'Permanently' should be understood in the context of the analysis which is the time period up until 2011. Moves made after 2011 are not recorded in the dataset analyzed in the current study.

Figure 6 reveals that (1) the largest percent of authors who stay in their country are US authors followed by Italian authors; (2) a much smaller percentage of authors move permanently and those are from predominantly from the UK, followed by Dutch and German authors, The least likely to move permanently are authors from Italy and USA. (3) Lastly, the percentage of authors who migrate and return is for each country represented in Figure 6 lower than that of authors moving permanently. The category "other" in Figure 6 contains researchers migrating to multiple destinations, for instance, back and forth between A and B several times.

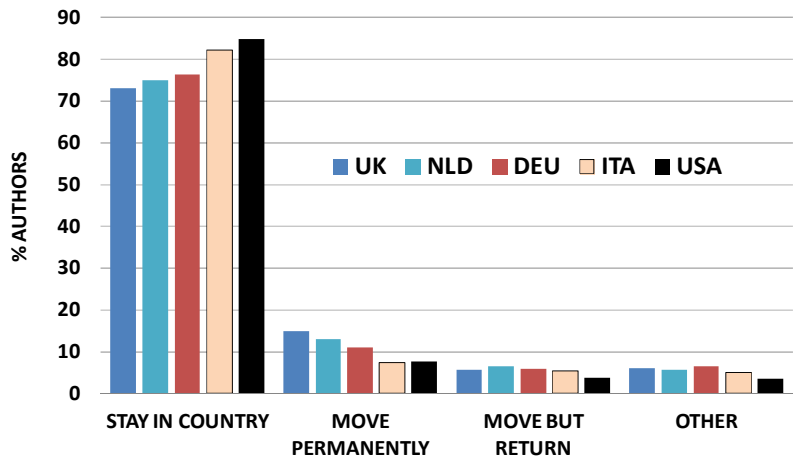

Figure 6: Asynchronous migration patterns for 5 countries.

It must be underlined that this ratio is expected to change with the length of the time period considered. The longer the time period during which the fixed cohorts are studied, the more information is available on them, and, hence, the larger is probability that a researcher's return to his home country is recorded in the data For instance, among the authors staying in the country in Figure 6, a part may move abroad after 2011; in addition, of the authors moving permanently in Figures 6 and 7, a certain fraction may return after 2011. Therefore, it is important to combine the asynchronous with the synchronous approach and to study sufficiently large time periods; it is their combination that provides a more complete view of migration patterns among countries.

The asynchronous approach follows authors who started their careers during 2001-2003. Figure 7 shows that the percentage of authors moving permanently abroad is larger than the percentage of authors that remain for all countries studied. But at the same time there are large differences among countries in the value of this ratio. A comparison of the percentages of authors who move permanently to those who move and return to their origin country, is shown in Figure 7. Considering the dashed linear regression line as a reference line, the figure shows that countries located upon or above the line tend to be scientifically developing, and those below the line scientifically developed, although there are a few exceptions. Thailand (THA), Malaysia (MYS), Portugal (PRT) and Egypt (EGY) are located in the graph substantially above the dashed regression line and thus seem to have a relatively large fraction of researchers who move abroad and come back. This could be the result of an intended strategy according to which researchers from these countries, after an initial research training, possibly a PhD stage, in their home country, seek to gain experience while studying abroad – possibly as post-docs – but return afterwards to their origin country to continue their careers, and contribute to building up a scientific infrastructure Researchers who moved to a foreign country and who decided to stay there may do so in quest for better professional or for a better economical and social life.

China (CHN) , Brazil (BRA), Iran (IRN)  and Australia (AUS) are positioned on or very near to the trend line, where the numbers are balanced between those who leave their country to work abroad and come back to those who leave permanently.

USA, Japan (JPN), UK but also India (IND)  are  positioned substantially below the regression line, where larger fractions  of researchers seem to be moving away to different countries permanently. This pattern may at least partly be caused by students from scientifically developing countries who move to these countries and start their scientific training there – possibly in a doctoral program –  and go back after attaining their PhD and continue their careers in the home country.  According to this interpretation, India plays in recent years the role of a scientifically developed country.

It must be underlined that Figure 7 only takes into account junior researchers who moved abroad from one of the 17  study countries

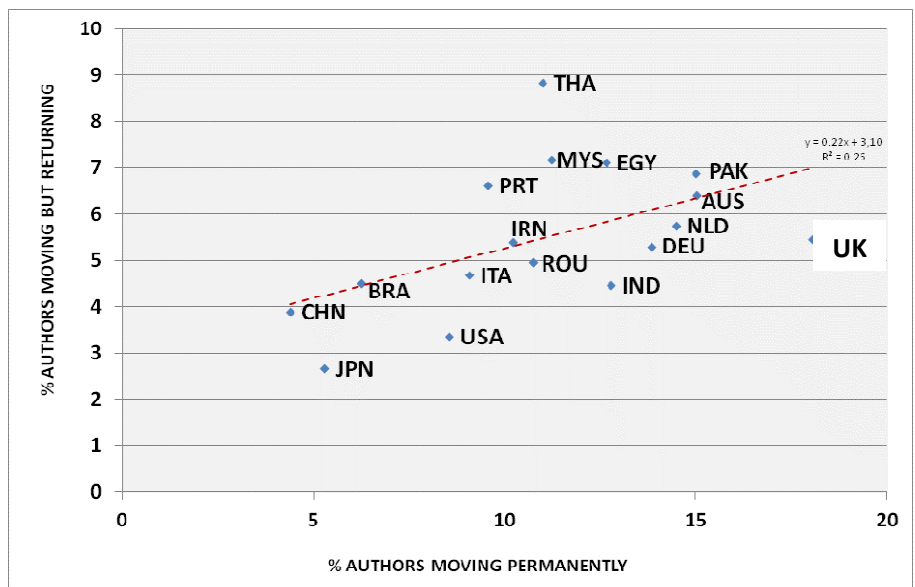

Figure 7: Migration patterns of study countries as source countries. The dashed line represents the linear regression line. For the corresponding full country names of the country codes the reader is referred to the legend to Figure 2.

## 5.  Accuracy of the migration indicators

In order to obtain more insight into the accuracy of the Scopus author profiling a case study was conducted analyzing a sample of 100 author IDs randomly selected from the subject field "Chemistry" in Scopus. These authors were active in over 80 different countries. The 6 countries with the largest number of authors in the sample were China, USA, Japan, Germany, Korea and India. Therefore, it can be concluded that Asian authors are somewhat overrepresented in the sample.

The analysis focused on recall rather than precision; It examined to which extent the researchers represented by the selected 100 author IDs were linked to other IDs in Scopus. In other words, to which extent the publication oeuvres of these researchers were split among two or more author IDs. These researchers will be denoted as unique authors below.

It was found that 27 per cent of the unique authors did have at least one additional author ID. For these authors 51 additional author IDs were found. The number of papers assigned to each of these additional author IDs was low: half of these had 1 paper only; and 75 per cent had 2 at most. A further analysis revealed that the number of author IDs with one single paper was two times the number of unique authors with one paper. This multiplication factor decreased rapidly with increasing number of articles assigned to an author ID or unique author; 1.6 for authors with 2 papers and 1.1 for authors with 20 or more papers. These outcomes make it necessary to further analyze the accuracy of the outcomes of the migration analyses presented in this paper.

Two decisions made in this paper reduced to some extent the error rates in the author profiling. First, authors publishing one single paper were defined as "transients". As outlined in Section 3, this category of authors was not included in the migration analysis. This decision substantially reduced the number of additional author IDs that should have been merged with other author IDs in the analyzed sets but were not properly captured by Scopus author profiling routine. Secondly, the analyses in this paper relate to author IDs that started publishing in 2001 or later. This decision eliminated a certain fraction of authors with common names – since truly common author names such as  Brown, Jones, or Liu can safely be assumed to appear as authors during the *entire* period 1996-2011, *not* only from 2001 onwards. Authors with such common names are not included in the analysis, thus reducing the error rate due to homonyms. However, it is difficult to specify the values of the two reduction rates.

The authors of this paper are not aware of the details of the algorithm underlying the current version of author profiling in Scopus. It must be noted that if an author's affiliation  does play a role in the algorithm, this may lead to structural bias in the migration data, to the extent that there would be a tendency that persons moving from one institution to another would be easily split into two author IDs.

An attempt was made to obtain at least a rough indication of the remaining error rates in the migration data in the following manner. This analysis focused on the following key migration-based indicators:  the percentage of authors starting their career in a study country, and moving abroad at least once; the percentage of authors moving from a study country to a particular other country; and the ratio of this percentage and the percentage of articles co-authored by authors from the two countries. The latter indicator compares international scientific migration and internatonal collaboration ties between two countries.

All indicators presented in this paper were re-calculated in two separate runs in which specific classes of authors based on their publication counts were eliminated. If the first run includes all author IDs regardless of the number of their published articles – the data presented in Section 4 are based on this run –, in a second run author IDs with 2 papers only were eliminated (in order to reduce the number of authors with a split identity), as well as those with 7 or more publications per year during 2001-2011 – in order to eliminate erroneous merging of publication oeuvres of different researchers into one author ID.

A third run eliminated authors with 2 or 3 articles, and those having more than 5 papers per year. In this way one can obtain some insight into the robustness of the results, and suggest plausible error rates for the various indicators. Figures 9 and 10 illustrate the effect of eliminating author IDs with low and high publication counts upon the indicators.

Figure 9 presents the percentage of authors starting their career in a study country during 2001-2003 and moving abroad at least once in the time period up until 2011, for the three author sets, and for each of the 17 study countries (the so called asynchronous approach for which Figures 6 and 7 present particular outcomes). Figure 9 shows a general tendency that the percentage of authors moving abroad increases if the authors with the lowest and largest publication counts are eliminated. This effect is mainly due to the elimination of author IDs publishing 2 or 3 papers only. Apparently migration in this set of author IDs is less frequent. It must be noted that eliminating these classes of authors, one does not only remove "wrong" author IDs that should have been merged with other author IDs in the total set, but also correct author IDs of researchers with a low publication count. Most importantly, this figure shows that there are very few cases in which the lines for two different countries intersect. It therefore illustrates the robustness of at least rankings among study countries based on the asynchronous migration indicator.

As a second example, Figure 10 shows the ranks of each pair of countries on the basis of the ratio of migration strength and co-authorship strength, calculated in the first (including all author IDs) and the second run which eliminates authors with 2 papers or more than 7 papers per year. It illustrates that the top of the ranking hardly changes if low and highly productive author IDs are eliminated. This reveals the robustness of the analysis comparing migration and co-authorship, at least for the top segment of the ranking.

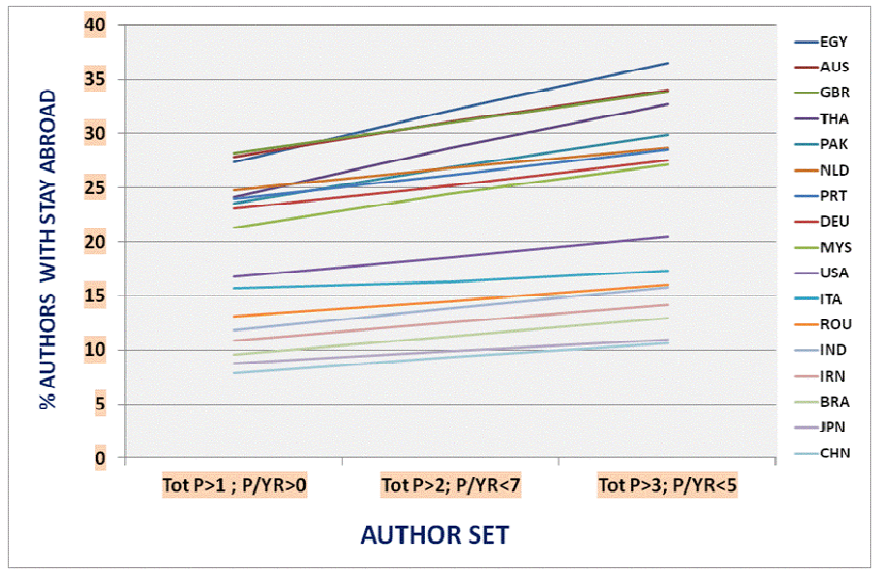

Figure 9: The effect of eliminating lowly and highly productive author ids upon the indicator: percentage of authors starting their career in a particular and moving abroad. For instance, Tot P>2; P/yr<7  means: the run is based on authors publishing more than 2 articles during 2001-2011, but less than an average of 7 articles per year.

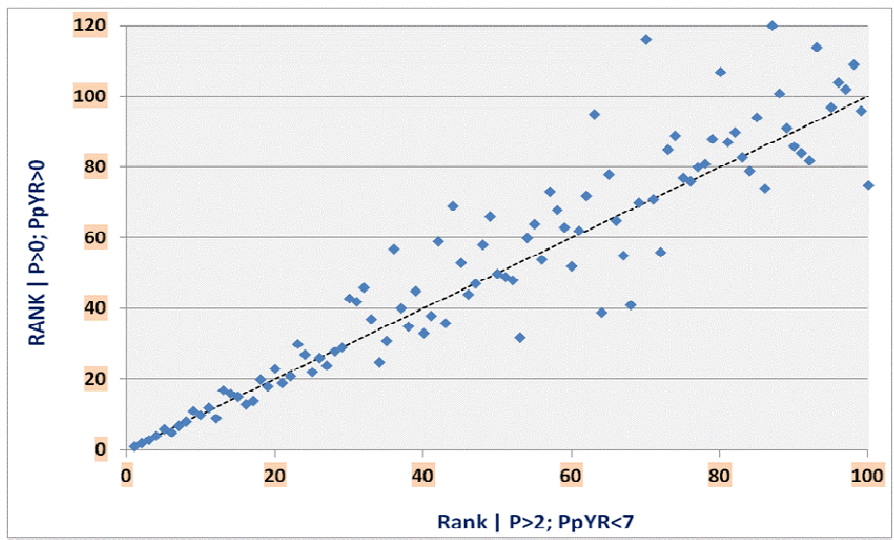

Figure 10: The effect of eliminating lowly and highly productive author ids upon the indicator: a country pair's ratio of migration and co-authorship. The vertical axis gives a country pair's rank according to its value of its ratio of migration and co- authorship (with pairs ranked by descending value of this ratio)  based on the set of all authors, while the horizontal axis gives these values based on authors publishing more than 2 but on average less than 7 papers during 2001-2011.

| Indicator | Set | N | Mean variation* | Standard Dev | Skewness | Rough error rate |
|-----------|-----|---|-----------------|--------------|----------|------------------|
|  |  |  |  |  |  |  |

| # or % publishing authors per country | All | 17 | 11.1 | 2.9 | -0.25 | 10 % |
|---|---|---|---|---|---|---|
| # or % migrating authors per country pair | Top 100 | 100 | 2.1 | 1.2 | 1.4 | 3 % |
| | All | 650 | 11.1 | 28.8 | 7.0 | >10 % |
| Ratio migration/collaboration per country pair | Top 100 | 100 | 2.8 | 2.4 | 2.4 | 5 % |
| | All | 650 | 10.6 | 27.6 | 7.4 | > 10 % |

Table 6: Rough error rates in migration indicators *Variation = 100*(max-min)/ (2*mean)

Table 6 presents tentative error rates for migration indicators. It compares for each unit of analysis (individual countries in the first indicator, and country pairs in the second and third) the values calculated for each of the three author sets. The variation is defined as the relative difference between the maximum and minimum value of the indicator. It readily increases as the unit's indicator vales are lower. For instance, as regards the indicator *# or % migrating authors per country pair* the mean variation over the top 100 of pairs is only 2.1 per cent. But for the subset of pairs ranking 101-200 it is 3.5, and for the subset of pairs ranking 501-600 it is almost 20 per cent. It can be concluded that in a ranking based on descending indicator values, countries in the uppor part of the raning show error rates of less than 10 per cent. But in lower parts of such a ranking the error may be larger than 10 per cent, and easily reach values of 20 per cent.

## 6. Discussion and conclusions

*General comments*

The results presented in this paper confirm the conclusion drawn in previous research (Laudel, 2003) that author affiliation data is in principle a valuable source of information in studies on international scientific migration. The use of large scientific literature database containing comprehensive information on authors and their affiliations, can potentially lead to new insights into the phenomenon of scientific migration, provided that the outcomes the bibliometric study are interpreted with care.

Using a bibliometric approach to analyze affiliations within articles and the ability to systematically attribute them to unique authors' profiles enables the study of migration trends. International scientific migration generates new insights into the global scientific network as it can potentially create a breeding pool for future international collaboration.

The bibliometric approach can be used not only descriptively to show for instance maps of international scientific migration traffic among countries, but also as a tool for studying the causes and effects of scientific migration. For instance, from the point of view of research assessment, an interesting issue is the relationship between migration on research performance,

A key limitation holds that the bibliometric methodology applied in this paper studies scientific migration through an analysis of scientific publications. If a researcher moves to a country but does not publish there for example, it will not show in the analysis. For migration trends to be tracked, the researcher has to publish at least once per year. This will allow an analyst to trace their scientific career. But if they publish occasionally or not at all, publication databases provide little if any information on researchers.

As outlined in the methodology section, a main limitation of author affiliation data is that they do not provide information on the country of birth of an author, or on the country in which he or she obtained a pre-doctoral degree. Even if one assumes that the country from which an author publishes his or her first article marks the national research system in which he/she starts his research training – which is statistically speaking a plausible assumption especially in fields with a strong publication culture – the analyst does not know where this author comes from: he could have attained the master degree in another country and moved to a PhD position in the country of his first paper. In other words, the purely affiliation-based approach may miss a relevant aspect of scientific migration.

On the other hand, the observation made above does not imply that the bibliometric approach to international scientific migration has no value at all. The current approach does give insight into how researchers starting their scientific career in a particular country, and continue this career, move from one country to another, either for temporary stays or permanent positions.

The results obtained in this exploratory paper indicate that the bibliometric approach based on author affiliation data is more appropriate to study migration taking place during the transition from the PhD-phase to the post-doctoral phase in a research career, than that from the master student phase to the PhD phase.

A second conclusion holds that the bibliometric approach would be more valuable if the author data would be enriched, for instance, by linking publishing authors to databases with CV information on researchers' education and nationality. The best approach to a comprehensive study of scientific migration is the combination of databases, especially publication and CV databases. Survey, questionnaire or country census data could also be useful complementary data. Bibliometric data could be used to validate and check the consistency of survey-based data.

*Data accuracy*

As stated above, this study is based on the use of unique authors' profiles compiled by Scopus™. The Scopus™ author profile algorithm assigns each author indexed in the database a unique identification number and aims to group each author's publications and affiliations. The algorithm used by Scopus™ to associate an author with his/her publication uses multifaceted approach where name variants, co-authors, subject areas and publications history is taken into account. The system aims at higher levels of precision rather than recall. Therefore, if the algorithm cannot determine whether a publication is indeed the product of a certain author, it will create an additional profile under which the publication will appear. However, it is important to note that the automatic author profiles generated by Scopus™ are supplemented by an author feedback system whereas an author can indicate whether there are publications that are missing from his/her profile or wrongly attributed to him/her.

There are, therefore, two scenarios that might affect the analysis in this study: (1) cases where an author unique profile might contain papers that in fact belong to another author. This could happen especially in cases of common names and the result is that the more common the name, more papers will be attributed to it. (2) An author body of work is split between different profiles due to inability to compile them. In this scenario, the first profile will contain most of the works but several smaller profiles will contain 1-3 papers typically. In our analysis the effect of this type of errors is reduced by the elimination of transients (authors publishing papers in one single year only). Also, the effect of common names is reduced by considering only authors who start their publication career in 2001 or later.

The sample of authors studied in the analysis of the author profiles in Scopus was drawn from a single discipline, chemistry. In our view this is not a severe limitation since it is not clear why author profiling errors would vary among disciplines. What is significant is that these problems can be expected to vary substantially among countries of origin, especially between Asian countries and Western countries. The sample studied is biased towards Asian countries, so it cannot be maintained that it may lead – or was aimed – to underestimating the errors in author profiling. More studies should be carried out on larger data samples.

The present case study sheds light on the possible effects of the inaccuracies in author profiling, at least on the outcomes of type of analyses presented in this paper. It suggests that error rates due to inaccuracies in author profiling for units of assessment with large indicator values – e.g., for the 25 countries with the largest ratio of migration /co-authorship – are fairly below 10 per cent. Although the actual values of the percentages of authors with a stay abroad change under the influence of errors in author profiles, rankings of countries according to this percentage are hardly affected.

*Further Research*

Understanding the motivation behind migration is important as it affects all of many aspects of society including science policy, economical competencies, politics and social trends. Therefore, future study is needed in order to examine the formation of research excellence centres and their attractiveness to international migration. A disciplinary analysis of migration could shed light on regional competencies and the manner by which they attract migration. This type of analysis has potential to influence science policy whereas pockets of regional disciplinary activity and levels of collaborations within them can be identified and better directed. If, for example, country A draws researchers from country B and C in the area of stem cell research for example, countries B and C might want to invest more in this area in order to retain their talents. Disciplinary analysis is also needed in order to better understand the relationship between co-authorships and migration. In this respect it is important to understand if co-authorship leads to collaboration or vice versa and to what extent. Findings of such study will enable in-depth look into the manner by which scientific networks forms, in what areas and will also enable the identification of sustainable networks as opposed to regional or occasional ones. Implications of this analysis could be better directive ability of research ventures and working relationships between countries.

## Acknowledgement

The authors wish to thank the useful comments made by anonymous referees.

## References

Abramo, G., D'Angelo, C. A., & Solazzi, M. (2012). A bibliometric tool to assess the regional dimension of university-industry research collaborations. *Scientometrics, 91*(3), 955-975.

Baruffaldi, S. H., & Landoni, P. (2012). Return mobility and scientific productivity of researchers working abroad: The role of home country linkages. *Research Policy, 41*(9), 1655-1665.

Biondo, A. E. (2012). What's up after brain drain? sometimes, somewhere, someone comes back: A general model of return migration. *International Review of Economics, 59*(3), 269-284.

Di Maria, C., & Lazarova, E. A. (2012). Migration, human capital formation, and growth: An empirical investigation. *World Development, 40*(5), 938-955.

Gutiérrez-Vela MM, Díaz-Haro A, Berbel-Salvador S, Lucero-Sánchez A, Robinson-García N, Cutando-Soriano A. (2012).Bibliometric analysis of research on regenerative periodontal surgery during the last 30 years. Journal *of Clinical and Experimental Dentistry, 4*(2), 112-118.

García, N., & Cutando-Soriano, A. (2012). Bibliometric analysis of research on regenerative periodontal surgery during the last 30 years. *Journal of Clinical and Experimental Dentistry, 4*(2), 112-118.

Laudel, G. (2003). Studying the brain drain: can bibliometric methods help?. *Scientometrics*, 57, 215–237.

Moed, H.F., Aisati, M., Plume, A. (2013), Studying scientific migration in Scopus. *Scientometrics* 94, 929-942.

Plume, A. (2012). The evolution of brain drain and its measurement: Part I. *Research Trends,* No. 26 (http://www.researchtrends.com/issue26-january-2012/the-evolution-of-brain-drain-and-its-measurement-part-i/).

Plume, A. (2012). The evolution of brain drain and its measurement: Part II, *Research Trends*, No. 27 http://www.researchtrends.com/issue-27-march-2012/the-evolution-of-brain-drain-and-its-measurement-part-ii/).

Snaith, B. (2012). Collaboration in radiography: A bibliometric analysis. *Radiography, 18*(4), 270-274.

Yang, C. C., & Tang, X. (2012). A content and social network approach of bibliometrics analysis across domains. Paper presented at the *ACM International Conference Proceeding Series,* 515-517.